\documentclass{article}

\usepackage{arxiv}
\usepackage{amsmath}
\usepackage[utf8]{inputenc} 
\usepackage[T1]{fontenc}    
\usepackage{hyperref}       
\usepackage{url}            
\usepackage{booktabs}       
\usepackage{amsfonts}       
\usepackage{nicefrac}       
\usepackage{microtype}      
\usepackage{lipsum}		
\usepackage{graphicx}
\usepackage{natbib}
\usepackage{doi}

\title{Playability-Aware Audio-to-Tablature Guitar Transcription via Diffusion Models}

\date{} 		
  
\author{ {\hspace{1mm}Riccardo Simionato} \\
	Univ. Bordeaux\\
	CNRS, Bordeaux INP, LaBRI\\
	Talence, France \\
	\texttt{riccardo.simionato@u-bordeaux.fr} \\
	\And
	{\hspace{1mm}Louis Bigo} \\
	Univ. Bordeaux\\
	CNRS, Bordeaux INP, LaBRI\\
	Talence, France \\
	\texttt{louis.bigo@u-bordeaux.fr} \\
}

\renewcommand{\shorttitle}{ }
\newcommand{\headeright}{ }
\newcommand{\undertitle}{ } 

\hypersetup{
pdftitle={A template for the arxiv style},
pdfsubject={q-bio.NC, q-bio.QM},
pdfauthor={David S.~Hippocampus, Elias D.~Striatum},
pdfkeywords={First keyword, Second keyword, More},
}

\begin{document}
\maketitle

\begin{abstract}
Guitar tablature transcription requires not only accurate pitch detection but also assigning each note to a specific string-fret position, as the same pitch can be played at multiple fretboard positions. Existing approaches treat this as a standard classification problem, ignoring the musical and physical constraints that govern playable fingering sequences. We propose Noise2Fret, a diffusion model for audio-to-tablature transcription that generates tablature through a continuous latent representation of discrete fret and string targets, conditioned on spectral and audio features. To bridge the gap between pitch accuracy and physical playability, we introduce five auxiliary losses encoding Pitch-Class Distance, Positional Distance, Circle-of-Fifths Distance, String Similarity, and Hand-Span Feasibility directly into the training objective. Experiments on GuitarSet and GOAT datasets demonstrate that the model outperforms baselines while remaining computationally more efficient, and that the auxiliary losses yield consistent gains over the standard training objective.
\end{abstract}

Automatic Music Transcription (AMT) has advanced substantially through deep learning, progressing from frame-level multi-pitch estimation \cite{app132111882} to full notation-level transcription across a hierarchy of frame, note, stream, and notation levels. At the frame level, the system predicts pitch activity within short time segments; at the note level, it detects individual notes with their onsets and offsets; at the stream level, it separates notes into distinct melodic lines or voices; and at the notation level, it generates full score notation, including rhythm and structural information. Notation-level transcription is particularly demanding, requiring not only pitch accuracy but also the encoding of instrument-specific information — fingering positions, playing techniques, articulations, and dynamics — making manual transcription laborious and motivating automatic systems.
\begin{figure}[h]
    \centering
    \includegraphics[width=0.3\linewidth, trim=0cm 0cm 0cm 4cm, clip]{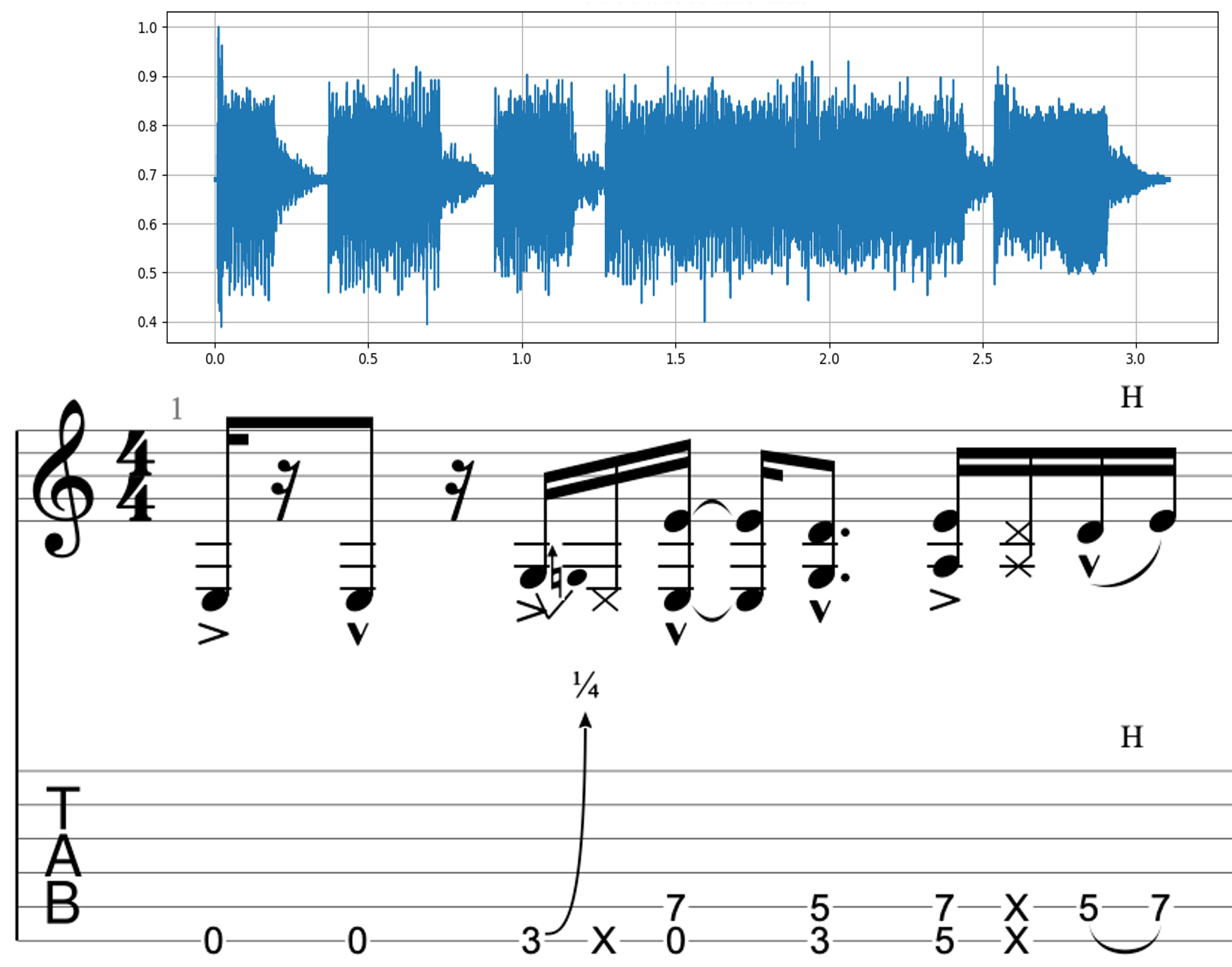}
    \caption{A guitar tablature (bottom) and its equivalent standard notation (top). The third note should be performed with a \emph{bend} and the last two with an \emph{hammer-on}.}
    \label{fig:tab}
\end{figure}
Guitar tablature transcription poses a distinct challenge: the guitar maps the same pitch to multiple string-fret combinations, requiring joint optimization of pitch estimation and fingering assignment \cite{chieppa2025automatic}. 

While playability has been partially addressed through post-hoc constraints \cite{yazawa2013audiotab} and losses that discourage unlikely note combinations \cite{cwitkowitz2022data}, these approaches encode what co-activations are statistically rare rather than why they violate physical or musical principles. The inhibition loss remains agnostic to hand-span geometry, fret-distance ordinality, and harmonic structure, and as a strictly pairwise model, cannot capture higher-order constraints involving three or more simultaneous fingers. No prior work has proposed a unified training objective jointly encoding physical feasibility and musical consistency within an end-to-end generative model.

Recent diffusion-based transcription models suggest a promising alternative to direct classification by iteratively refining discrete musical representations in a learned continuous space. In particular, Diﬀroll \cite{cheuk2023diffroll} introduced diffusion-based generative music transcription with unsupervised pretraining, and D3RM \cite{kim2025d3rm} showed that discrete denoising diffusion can be effective for piano transcription, motivating the use of a diffusion formulation for guitar tablature as well.

In this paper, we propose Noise2Fret, a novel architecture and loss functions for audio-to-tablature transcription that directly address this gap. Our contributions are:
\begin{itemize}
    \item Five auxiliary losses encoding physical feasibility and musical consistency into the training objective.
    \item A low-latency and efficient architecture that models tablature as a sequence of temporally ordered events, explicitly capturing inter-note dependencies within a generative framework.
\end{itemize}

The paper is organized as follows. Section \ref{sec:related} presents the state-of-the-art, Section \ref{sec:representation} details the data representation, and Section \ref{sec:methods} explains methods, neural architectures, losses, and evaluation metrics designed for the task. Section \ref{sec:results} discusses the results, and finally, Section \ref{sec:conclusion} concludes the paper. All the source code for this paper, along with trained models, is available online\footnote{\url{https://github.com/RiccardoVib/Noise2Fret}}.

\section{Related Work}
\label{sec:related}

Early work on music transcription relied on signal processing and Non-negative Matrix Factorization (NMF)-based methods \cite{smaragdis2003non, vincent2009adaptive, benetos2013multiple}, while modern systems increasingly exploit CNNs and Transformer architectures. The Onsets and Frames model \cite{hawthorne2017onsets} established the dual-objective note detection paradigm for piano, later generalized to sequence-to-sequence token decoding \cite{hawthorne2021sequence} and lightweight convolutional inference \cite{bittner2022lightweight}. Multi-instrument transcription has been further addressed by MT3 and its successors \cite{gardner2021mt3, tan2024mr, chang2024yourmt3}.

Guitar transcription research also relied on traditional signal processing methods. One of the first comprehensive systems for electric guitar transcription jointly estimated score-related parameters — note onset, duration, and pitch — alongside instrument-specific information such as string number, plucking style, and expressive techniques \cite{kehling2014automatic}. The incorporation of playability constraints was also explored by modeling physically feasible fingering configurations via dynamic programming \cite{yazawa2013audiotab}.

TabCNN \cite{wiggins2019guitar} pioneered end-to-end audio-to-tablature mapping using CNNs trained on GuitarSet \cite{xi2018guitarset}, demonstrating that spectral-temporal features implicitly encode playability constraints. Subsequent works introduced self-attention and beat-informed quantization, which quantize latent features corresponding to the given BPM
information \cite{kim2022note}, and large-scale synthesized datasets to mitigate data scarcity \cite{zang2024synthtab}. Playability constraints have also been incorporated by estimating the pairwise likelihood of co-occurring string-fret combinations from tablature \cite{cwitkowitz2022data}. Finally, FretNet \cite{cwitkowitz2023fretnet} jointly estimates continuous-valued pitch contours and string/fret assignments from polyphonic guitar audio. 

Chen et al. \cite{chen2022towards} broadened electric guitar transcription research by introducing the EGDB dataset and a multi-loss Transformer model, demonstrating both the importance of task-specific objectives and the sensitivity of transcription performance to timbre. Riley et al. \cite{xavier2024high} later showed that high-resolution domain adaptation from piano transcription can yield strong zero-shot guitar transcription results on GuitarSet. More recently,  Riley et al. \cite{riley2024gaps} introduced GAPS, a large real-recorded classical guitar dataset and benchmark model, further emphasizing the value of diverse, realistic data for generalizable transcription. Chieppa et al. \cite{chieppa2025automatic} reused beat-informed quantization for guitar transcription, reinforcing the importance of temporal structure and note-level prediction. Importantly, these works mainly predict note-level attributes such as pitch and onset/offset, rather than explicit string–fret assignments.

\section{Data representation}
\label{sec:representation}

\subsection{Discrete to Continuous Tablature Representation}

Guitar tablature is a symbolic notation system, as can be seen in Figure \ref{fig:tab}, that encodes guitar performances as a sequence of fret and string assignments. It captures not only pitch but also the physical fingering position, together with playing techniques such as bends, hammer-on/pull-off, slides, or vibratos. In this work, tablature data is sourced from Guitar Pro files (.gp) \cite{dadagp2021}, a widely used format for digital score editing that stores note events with string, fret, onset, duration, and playing technique annotations, and JAMS (JSON Annotated Music Specification) format, which encodes note-level transcriptions including string and fret assignments alongside beat, chord, and tempo metadata. These symbolic representations provide exact ground-truth fingering information that standard MIDI cannot supply.

The system frames audio-to-tablature transcription as a discrete process over event sequences, where an event denotes a set of one or several strings sounding simultaneously. Each event is represented as a sequence of $6$ per-string class labels, where each label encodes one of $F$ possible states: muted, open, or frets $[1, M_f]$, with $M_f$ the maximum fret position in the dataset and resulting in $F = M_f + 2$. 

We project the discrete tablature vector into a continuous embedding space. For each string, we use a learned embedding table that maps the $F$ fret classes to an $E$-dimensional vector, and then concatenate the $6$ embeddings into a single per-event representation of dimension.

\subsection{Audio Input Features}

The design method exploits a set of features $\mathcal{F}$, which consists of Short-time frequency transform (\textit{STFT}) magnitude, spectral flux \cite{dixon2006spectralflux} (\textit{SF}), and brightness \cite{grey1978perceptual} (\textit{B}), extracted from $16$ kHz mono waveforms input audio $\mathcal{A}$. 

\textit{STFT} magnitudes serve as the primary pitch representation and are extracted using a $1024$-point FFT with a $256$-sample hop and a Hann window, producing $T$ STFT frames with $513$ frequency bins each.

\textit{SF} measures the rate of spectral change between consecutive STFT frames via the half-wave rectified $\mathbf{L}_2$ norm, yielding an onset strength signal of $T$ frames. Sharp increases in spectral energy correspond to attack transients, providing the model with explicit temporal anchors for note boundaries.

\textit{B} encodes the distribution of spectral energy and helps the model disambiguate physically equivalent pitches across different string-fret positions: higher fret positions increase string stiffness, damping high-frequency overtones, and lowering the spectral centroid. Defined as the ratio of energy above $1,500$ Hz to total frame energy, it provides a frame-level signal of shape $(T \times 1)$. 

\section{Methodology}
\label{sec:methods}

\begin{figure}
    \centering
    \includegraphics[width=0.7\linewidth, trim=0cm 0cm 0cm 0cm, clip]{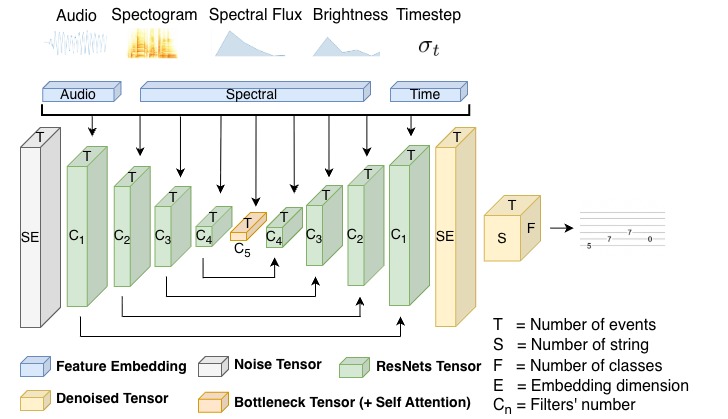}
    \caption{Overview of the proposed Noise2Fret architecture at inference time. Starting from a Gaussian noise tensor in the continuous embedding space ($T \times SE$), the model iteratively denoises the representation through a 1D convolutional U-Net comprising four encoder stages, a self-attention bottleneck, and a symmetric decoder with skip connections. Audio, spectral features, and timestep are injected as conditioning signals at each resolution level. The final denoised embedding is projected back to per-string class logits over $F$ fret states, yielding the predicted tablature tensor ($T \times S \times F$).}
    \label{fig:model}
\end{figure}

Figure~\ref{fig:model} illustrates the Noise2Fret architecture. The input tablature is embedded into a continuous space, corrupted with Gaussian noise, and denoised by a 1D convolutional U-Net with three encoder–decoder stages and a self-attention bottleneck, conditioned on raw audio, spectral magnitude, spectral flux, and brightness. 
The decoder outputs are projected to per-string fret logits over $F$ fret classes, which include muted and open strings, and the final prediction is obtained by selecting the highest-scoring class.

\subsection{Diffusion Framework}

Diffusion models learn to reverse a gradual noising process, transforming structured data into noise during training and recovering it during inference. We adopt a continuous diffusion framework operating in the learned tablature embedding space, following the v-prediction parameterization of \cite{salimans2022progressive}.

The forward process corrupts the target tablature embedding $x_0$ by interpolating with Gaussian noise $\epsilon \sim \mathcal{N}(0, I)$ according to a cosine noise schedule $x_t = \alpha_t x_0 + \beta_t \epsilon$, where $\sigma_t \in $ controls the noise level at step $t$ and $ \alpha_t = \cos\!\left(\sigma_t \frac{\pi}{2}\right)$, $\beta_t = \sin\!\left(\sigma_t \frac{\pi}{2}\right)$.

Rather than predicting the noise $\epsilon$ directly, the model is trained to predict the velocity target $v = \alpha_t \epsilon - \beta_t x_0$, from which a clean estimate $\hat{x}_0$ is recovered as $\hat{x}_0 = \alpha_t x_t - \beta_t v$. 

The model is conditioned on audio features extracted from the input recording, such that the denoising process is guided toward the tablature consistent with the observed performance. During training, timesteps $\sigma_t$ are sampled uniformly from $[0, 1]$.

At inference, denoising is performed iteratively over a linear sigma schedule using a deterministic DDIM \cite{song2020denoising} sampling strategy, which minimizes the number of required function evaluations while maintaining transcription quality.

\subsection{Architecture Design}

The denoising backbone is a 1D convolutional U-Net operating along the sequence dimension. The input is first projected to the base channel width $C$, and sinusoidal positional embeddings are added to the noisy token sequence. The encoder consists of four \textit{ResNet} blocks with channel widths of $C$, $2C$, and $4C$, followed by a strided convolution that further downsamples the representation while keeping $8C$ channels. A middle \textit{ResNet} block operates at $8C$ channels and includes self-attention.

The decoder mirrors this structure with a transposed-convolution upsampling stage and four \textit{ResNet} blocks, using skip connections that concatenate encoder and decoder activations at matching resolutions. A final normalization, activation, and convolutional projection maps the hidden representation back to the per-string class logits.

Each \textit{ResNet} block contains two convolutional sublayers with GroupNorm and SiLU activations, a residual connection, diffusion-timestep embedding conditioning through a learned linear projection, diffusion-timestep embedding conditioning through feature modulation via FiLM \cite{perez2018film}, and feature injection from the spectral and audio conditioning streams. Conditioning feature injection is performed through a dedicated \textit{Injection Block}, which linearly projects the spectral or audio feature map to the current channel width with a $1 \times 1$ convolution, interpolates it to the target sequence length when necessary, and adds it residually to the block activations.

\subsection{Conditioning Strategy}

The model receives the conditioning signals at every \textit{ResNet} block: spectral features (\textit{STFT}, \textit{SF}, and \textit{B}), raw audio ($\mathcal{A}$), and diffusion timestep ($\sigma_t$). The spectral features are projected with a $1 \times 1$ convolution to the block width and injected at each block, while $\mathcal{A}$ is encoded by a three-layer strided CNN into a $64$-dimensional embedding sequence of length $\lfloor L/64 \rfloor$, with $L$ the input length in samples. This embedding is also injected independently at each block alongside the spectral features. 

Together, these signals provide the model with a multi-faceted view of the input: pitch structure (\textit{STFT}), temporal boundaries (\textit{SF}), playing position (\textit{B}), and low-level acoustic detail ($\mathcal{A}$). $\mathcal{A}$ is included as a complementary conditioning signal, preserving fine-grained temporal and timbral details. Finally, $\sigma_t$ is encoded via Gaussian Fourier projection \cite{tancik2020fourier}, producing a $128$-dimensional time embedding, which is injected and used via FiLM.

\subsection{Musical Auxiliary Losses}

We propose a suite of musically and physically informed loss functions that encode guitar constraints—spanning playability, harmony, and tonal structure—directly into the training objective, namely Pitch-Class Distance, Circle of Fifth Distance, Positional Distance, String Similarity, and Hand Span Feasibility loss.

The Pitch-Class Distance Loss is defined as the Jaccard distance between pitch-class sets:
\begin{equation}
    \mathcal{L}_{\text{pc}}(A, B) = 1 - \frac{|A \cap B|}{|A \cup B|}
\end{equation}
where $A$ and $B$ are the pitch-class sets of the predicted and target note events, respectively. This loss penalizes predictions whose active pitch classes diverge from the ground truth. In practice, each note event is mapped to a 12-dimensional binary pitch-class vector, with a value of 1 at each index corresponding to a pitch class present in the event, and the Jaccard distance is computed between the predicted and target vectors.

The Circle of Fifths Distance Loss penalizes predictions that are harmonically distant from the target at the tonal level. Each of the $12$ pitch classes is assigned a position $c(i)$ on the Circle of Fifths (CoF), where each step corresponds to a perfect fifth (7 semitones). A chord is represented by the weighted mean CoF position of its binary pitch-class vector $\mathbf{PC} \in \{0,1\}^{12}$: 
\begin{equation}
\mathcal{P}({\text{PC}}) = \frac{\sum_{i}^{11} \text{PC}_i \cdot c(i)}{\sum_{i} \text{PC}_i}
\end{equation}
The harmonic distance between two chords is then defined as the circular distance between their mean CoF positions:
\begin{equation}
\mathcal{L}_{\text{CoF}}(p, q) = \min\left(|\mathcal{P}(p) - \mathcal{P}(q)|,\ 12 - |\mathcal{P}(p) - \mathcal{P}(q)|\right)
\end{equation}
where $p$ and $q$ are the binary pitch-class vectors of reference and predicted chords. This loss is complementary to the Pitch-Class Distance Loss: while the latter penalizes pitch-class mismatches by treating all incorrect pitch classes equally, $\mathcal{L}_{\text{CoF}}$ penalizes predictions whose active pitch classes are tonally distant from the target, weighting errors by their harmonic distance on the Circle of Fifths.

The Positional Distance Loss measures the physical hand position. It is computed as the mean absolute fret distance between predicted and target fret assignments, excluding open strings and muted string slots:
\begin{equation}
\mathcal{L}_{\text{pos}} = \frac{1}{|S|}\sum_{s \in S} \left| f_s^{\text{pred}} - f_s^{\text{target}} \right|
\end{equation}
where $S$ denotes the set of active, non-open strings and $f_s$ is the fret position assigned to string $s$.

The String Similarity Loss measures the overlap between predicted and target active strings using the Jaccard distance. It follows the same formulation as $\mathcal{L}_{\text{pc}}$, but operates on a 6-dimensional binary vector $\mathbf{s} \in \{0,1\}^6$ where each element indicates whether the corresponding string is active. Therefore, $\mathcal{L}_{\text{str}}$ directly penalizes incorrect string selection.

The Hand Span Feasibility Loss penalizes fingering predictions that exceed a physically feasible hand stretch. It is defined as the normalised excess span beyond the maximum allowable fret distance $M_{\text{span}}$:
\begin{equation}
\mathcal{L}_{\text{span}} = \max\left(0,\ \frac{(f_{\max} - f_{\min}) - M_{\text{span}}}{M_{\text{span}}}\right)
\end{equation}\label{eq:span}
where $f_{\max}$ and $f_{\min}$ are the maximum and minimum fret positions among concurrently fretted, non-open strings, and $M_{\text{span}}$ is the maximum physically feasible hand span. The $\max(0, \cdot)$ hinge ensures the loss is zero for playable configurations and penalises only predictions that require an implausible stretch.

The auxiliary losses are designed to remain differentiable by operating on soft predictions, such as probability distributions over fret or pitch-class choices, instead of hard one-hot tablature labels. 

\subsection{Evaluation}

Performances are assessed using the standard AMT tablature metrics of \cite{wiggins2019guitar}: Pitch Precision (PP), Pitch Recall (PR), Pitch F-measure (PF), Tab Precision (TP), Tab Recall (TR), Tab F-measure (TF), and Tab Disambiguation Rate (TDR). All metrics are computed at the note level by matching predicted and ground-truth notes within a window around their onsets. Since the model predicts multiple notes per iteration, the final transcription is obtained by aggregating the notes across all iterations.
PP and PR measure the fraction of predicted and ground-truth pitches that are correct, respectively, with PF as their harmonic mean. TP and TR mirror these at the string-fret level, and TDR $= \text{TP} / \text{PP}$ quantifies how well the model resolves a correctly detected pitch to the right string and fret position.

We further introduce two complementary metrics: the False Negative Rate (FNR), which quantifies the proportion of played notes the model fails to predict, and the False Positive Rate (FPR), which quantifies the proportion of silent strings incorrectly predicted as active. The FNR is computed over target active string slots only, whereas the FPR is computed over target muted string slots only. 

A high FNR coupled with a low FPR indicates that the model is overly conservative, systematically preferring muted predictions over committing to a fret assignment.

Performances are evaluated against models that address the same problem formulation and output representation: the prediction of strings/frets assignments, such as TabCNN \cite{wiggins2019guitar} and FretNet \cite{cwitkowitz2023fretnet}. TabCNN has been re-implemented in Torch, and FretNet has been adapted to be compatible with the GOAT dataset.

\section{Experiments}
\label{sec:setup}

\subsection{Datasets}

In this study, we employed two datasets: GuitarSet \cite{xi2018guitarset} and GOAT \cite{loth2025goat}. GuitarSet comprises approximately $3$ hours of annotated acoustic guitar recordings, with string- and fret-level transcriptions provided in JAMS format alongside tempo, key, beat, and chord annotations. To ensure a fair comparison with prior work, we follow the same data preprocessing as in \cite{cwitkowitz2023fretnet} and employ a six-fold cross-validation scheme. 

The GOAT dataset contains $5.9$ hours of real guitar recordings paired with Guitar Pro tablature annotations covering fret/string numbers and expressive playing techniques. For training, we align audio excerpts to annotated onsets and crop a window starting at each onset; a single window may contain multiple notes or chords, and the target is the full tablature content in that window.

Experiments are restricted to standard-tuned recordings without effects processing; non-standard tunings, electric timbres, and capo-containing examples are excluded. The audio window is $100$ ms-long, sliced into $T = 7$ frames, and supporting up to $2$ events. Finally, the dataset consisted of $45,972$ examples, and the GOAT original train/validation/test splits are used for partitioning.

\subsection{Training Details}

Following Diffusion-LM \cite{li2022diffusion}, we perform the diffusion process in a continuous embedding space over discrete tablature. Logits over the $F$ classes are recovered by projecting the predicted embedding against the learned class vectors. The training objective combines an MSE loss in embedding space with a cross-entropy rounding loss weighted at $\lambda = 0.1$.

The total training objective combines the base diffusion loss with the auxiliary terms: $\mathcal{L} = \mathcal{L}_{\text{diff}} + \mathcal{L}_{\text{aux}}$, where
\begin{align}
\mathcal{L}_{\text{aux}} &= 
\lambda_{\text{pos}}\mathcal{L}_{\text{pos}} +\lambda_{\text{pc}}\mathcal{L}_{\text{pc}} \nonumber \\
& + \lambda_{\text{CoF}}\mathcal{L}_{\text{CoF}} + 
\lambda_{\text{str}}\mathcal{L}_{\text{str}} + \lambda_{\text{span}}\mathcal{L}_{\text{span}}.
\end{align}
The weights $\lambda_{\text{pos}} = 1$, $\lambda_{\text{pc}} = 0.1$, $\lambda_{\text{CoF}} = 0.1$, $\lambda_{\text{str}} = 0.1$, and $\lambda_{\text{span}} = 0.1$ are calibrated to maintain comparable loss magnitudes across all terms. The auxiliary losses are partially redundant with the base diffusion loss, but they are intentionally introduced to sharpen the penalty on musically meaningful errors that a plain MSE objective treats uniformly. Since GuitarSet contains performances with a capo, $\lambda_{\text{span}}$ is set to $0$ for this dataset to prevent the Hand Span Loss from incorrectly penalizing feasible fingerings.

All our models use a base channel width of $64$ and are trained for up to $1000$ epochs with AdamW \cite{loshchilov2019decoupled} at an initial learning rate of $3 \times 10^{-4}$, decayed with cosine annealing \cite{loshchilov2017sgdr} and a batch size of $128$. The hand-span threshold $M_\text{span}$ is set to $6$ frets, the maximum span observed in GOAT. 

To isolate the contribution of each design choice, we conduct a systematic ablation study in which individual components are removed one at a time from the full model. We consider two axes of ablation: (1) the auxiliary loss functions and (2) the conditioning signals.

\begin{table*}[!h]
  \centering
  \begin{tabular}{|l|ccccccc|}
    \hline
   Model & PP $\uparrow$ & PR $\uparrow$ & PF $\uparrow$ & TP $\uparrow$ & TR $\uparrow$ & TF $\uparrow$ & TDR $\uparrow$\\
    \hline
    Noise2Fret & $0.852$ & $0.854$ & $0.853$ & $0.841$ & $0.842$ &  $0.841$ & $0.987$ \\
     Noise2Fret (+$\mathcal{L}_\text{aux}$) & 0.855 & $\mathbf{0.857}$ & $\mathbf{0.856}$ & $\mathbf{0.844}$ & $\mathbf{0.846}$ & $\mathbf{0.845}$ & $\mathbf{0.988}$ \\
    TabCNN \cite{wiggins2019guitar} & $0.902$ & $0.759$ & $0.820$ & $0.776$ & $0.673$ & $0.717$ & $0.860$\\
    FretNet \cite{cwitkowitz2023fretnet} & $\mathbf{0.919}$ & $0.742$ & $0.818$ & $0.801$ & $0.669$ & $0.727$ & $0.871$\\
    \hline
  \end{tabular}
  \caption{Performance metrics for proposed and baseline models when trained and evaluated on the GuitarSet dataset.}
  \label{tab:GuitarSet}
\end{table*}

\section{Results}
\label{sec:results}

Table~\ref{tab:GuitarSet} reports performance on GuitarSet. The proposed $64$-dimensional model outperforms baselines across all F-measure metrics, demonstrating that the diffusion-based sequence approach generalises better than both TabCNN and FretNet. The addition of the auxiliary losses slightly improves performance across all metrics, confirming that explicitly encoding physical and musical constraints into the training objective yields consistent gains. 

On the other hand, pitch precision is marginally lower than the baselines, suggesting a slight trade-off outweighed by the substantial recall improvement, indicating the model recovers a greater proportion of ground-truth notes. The near-perfect Tab Disambiguation Rate confirms that the model reliably resolves string-fret ambiguity once a pitch is correctly detected.

Table~\ref{tab:loss} reports the metrics for the GOAT dataset along with the ablation study over the auxiliary loss components. The absolute metric values are lower than those observed on GuitarSet, reflecting the increased difficulty of the GOAT benchmark, which contains more complex fingering patterns and a wider variety of playing styles. The proposed model performed better across all metrics, although the baselines were re-implemented and adapted to GOAT, which may slightly affect their performance.  

Among individual losses, $\mathcal{L}_\text{pos}$ achieves the highest recall and the lowest FNR, while $\mathcal{L}_\text{pc}$ preserves high precision and a low FPR at the expense of recall. In contrast, $\mathcal{L}_\text{CoF}$ shows the most pronounced degradation, suggesting Circle-of-Fifths proximity is a weaker standalone inductive bias. $\mathcal{L}_\text{str}$ and $\mathcal{L}_\text{span}$ act more as regularisers, yielding a more balanced precision–recall trade-off with $\mathcal{L}_\text{str}$ leading to moderate improvements in TDR. TDR remains consistently high across all configurations, confirming that errors are predominantly missed detections rather than string-fret misassignments. The combined loss $\mathcal{L}_\text{aux}$ achieves consistent gains across almost all metrics, while maintaining competitive recall, indicating that the five components provide useful inductive biases and that their individual precision–recall trade-offs are resolved when jointly optimised.
\begin{table*}[h!]
  \centering
  \begin{tabular}{|l|ccccccccc|}
    \hline
   Model & PP $\uparrow$ & PR $\uparrow$ & PF $\uparrow$ & TP $\uparrow$ & TR $\uparrow$ & TF $\uparrow$ & TDR $\uparrow$ & FNR $\downarrow$ & FPR $\downarrow$\\
    \hline
    TabCNN \cite{wiggins2019guitar} & $ 0.753$ & $ 0.594$ & $ 0.664$ & $0.744 $ & $ 0.587$ & $0.656 $ & $0.987$ & $0.293$ & $0.047$\\ 
    FretNet \cite{cwitkowitz2023fretnet} & $0.689$ & $0.650$ & $0.669$ & $0.681$ & $0.643$ & $0.662$ & $0.989$ & $0.171$ & $0.066$\\
    Noise2Fret & $0.765$ & $0.729$ & $0.747$ & $0.757$ & $0.724$ & $0.740$ & $0.990$ & $\mathbf{0.145}$ & $0.031$\\
        \hline
    Noise2Fret (+$\mathcal{L}_\text{pos}$)  & $0.795$ & $\mathbf{0.791}$ & $0.793$ & $0.787$ & $\mathbf{0.785}$ & $0.786$ & $0.989$ & $\mathbf{0.145}$ & $0.031$ \\
    Noise2Fret (+$\mathcal{L}_\text{pc}$) &  $0.813$ & $0.721$ & $0.764$ & $0.804$ & $0.715$ & $0.757$ & $0.989$ & $0.216$ & $0.023$ \\
    Noise2Fret (+$\mathcal{L}_\text{CoF}$) & $0.722$ & $0.746$ & $0.734$ & $0.709$ & $0.735$ & $0.722$ & $0.981$ & $0.146$ & $0.040$  \\
    Noise2Fret (+$\mathcal{L}_\text{str}$) &   $0.792$ & $0.763$ & $0.777$ & 0.785 & $0.758$ & $0.772$ & $0.992$ & $0.163$ & $0.028$ \\
    Noise2Fret (+$\mathcal{L}_\text{span}$) & $0.805$ & $0.763$ & $0.783$ & $0.793$ & $0.754$ & $0.773$ & $0.985$ & $0.177$ & $0.028$ \\
    Noise2Fret (+$\mathcal{L}_\text{aux}$)  & $\mathbf{0.828}$ & $0.774$ & $\mathbf{0.800}$  &  $\mathbf{0.822}$ & $0.769$ & $\mathbf{0.795}$  & $\mathbf{0.993}$ &  $0.165$ &  $\mathbf{0.022}$\\ 
    \hline
  \end{tabular}
  \caption{Performance metrics for proposed and baseline models when trained and evaluated on the GOAT dataset (Top), and ablation study on designed auxiliary losses for the Noise2Fret model (Bottom).}
  \label{tab:loss}
\end{table*}

\begin{figure}
\centering
    \includegraphics[width=0.7\linewidth]{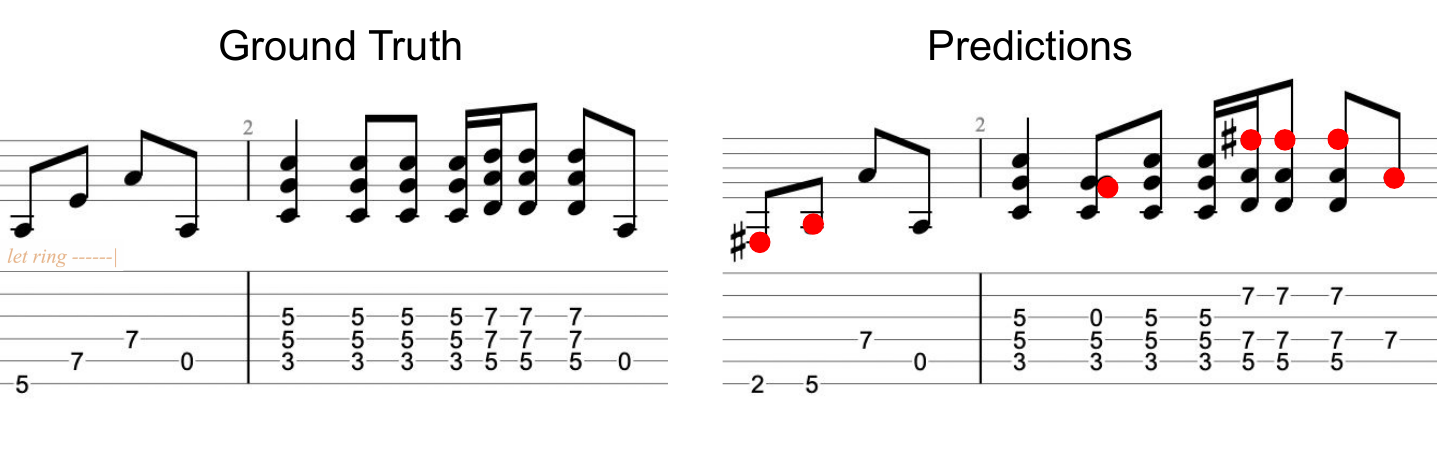}
    \caption{Example from the GOAT test set. Ground-truth tablature (left) and Noise2Fret prediction (right), with red markers indicating incorrect string-fret assignments. This example also illustrates a case where the ground-truth tablature lacks \emph{Let Ring} annotations, added in orange for reference.}
    \label{fig:example}
\end{figure}

Table~\ref{tab:ablations} reports the conditioning modality ablation. The full configuration achieves the best performance across all metrics, confirming that each modality contributes complementary information. Removing STFT causes the largest degradation, while removing brightness yields a modest drop in precision, and removing spectral flux affects recall more strongly. Ablating raw audio also lowers recall and increases FNR, showing that the waveform contributes low-level information not fully replaced by spectral features. TF and PF remain stable across configurations including STFT, indicating that the model preserves overall pitch-level transcription quality even when individual conditioning cues are removed. The TDR also remains high, suggesting that most errors are missed or incorrect pitch detections rather than string-fret misassignments, as shown in Figure~\ref{fig:example}. This suggests that the removed features influence pitch detection more strongly than string-fret selection.
\begin{table*}[!h]
  \centering
  \begin{tabular}{|l|ccccccccc|}
    \hline
   Conditioning & PP $\uparrow$ & PR $\uparrow$ & PF $\uparrow$ & TP $\uparrow$ & TR $\uparrow$ & TF $\uparrow$ & TDR $\uparrow$ & FNR $\downarrow$ & FPR $\downarrow$\\
    \hline
    \textit{STFT}+\textit{SF}+\textit{B}+$\mathcal{A}$ & $\mathbf{0.828}$ & $\mathbf{0.774}$ & $\mathbf{0.800}$  &  $\mathbf{0.822}$ & $\mathbf{0.769}$ & $\mathbf{0.795}$  & $\mathbf{0.993}$ &  $\mathbf{0.165}$ &  $\mathbf{0.022}$ \\
    \textit{SF}+\textit{B}+$\mathcal{A}$ & $0.227$ & $0.210$ & $0.219$ & $0.163$ & $0.151$ & $0.157$ & $0.715$ & $0.367$ & $0.066$ \\
    \textit{STFT}+\textit{SF}+$\mathcal{A}$ & $0.787$ & $0.756$ & $0.771$ & $0.776$ & $0.748$ & $0.761$ & $0.985$ & $0.168$ & $0.029$\\
    \textit{STFT}+\textit{B}+$\mathcal{A}$ & $0.796$ & $0.749$ & $0.772$ & $0.785$ & $0.742$ & $0.763$ & $0.986$ & $0.169$ & $0.025$ \\
    \textit{STFT}+\textit{SF}+\textit{B} & $0.796$ & $0.747$ & $0.771$ & $0.787$ & $0.741$ & $0.763$ & $0.988$ & $0.182$ & $0.027$ \\
    \hline
  \end{tabular}
  \caption{Ablation study on conditioning features for the Noise2Fret model trained on the GOAT dataset: Short-time frequency transform (\textit{STFT}) magnitude, spectral flux (\textit{SF}), brightness  (\textit{B}), and $16$ kHz mono waveforms as input audio $\mathcal{A}$.}
  \label{tab:ablations}
\end{table*}

Lastly, Noise2Fret presents $15$M trainable parameters, making it larger than TabCNN ($834$K parameters) and FretNet ($8.4$M parameters). However, in terms of floating-point operations per second (FLOPS), Noise2Fret is significantly more efficient than both baselines, requiring only $341$M operations per forward pass compared to $3.4$B for TabCNN and $17.4$B for FretNet, and achieving a Real-time Factor of $0.67$ on a single NVIDIA GeForce RTX 3090 GPU.

\section{Conclusion}
\label{sec:conclusion}

This paper introduces Noise2Fret, a diffusion-based model for automatic guitar tablature transcription that jointly addresses pitch detection and string-fret assignment in an end-to-end generative framework. Experiments demonstrate superior performance to TabCNN and FretNet on F-measure metrics. Five auxiliary losses are proposed to encode physical feasibility and musical consistency directly in the training objective. Although the individual losses involve precision–recall trade-offs, their combination yields broadly improved performance across the evaluated metrics.
The conditioning ablation shows that audio features add useful complementary information. Combined with a favourable computational profile, Noise2Fret represents a practical and principled step towards deployable guitar tablature transcription.

Future work will extend the system to playing techniques, non-standard tunings, and investigate architectural optimisations for real-time transcription. It will also examine missing annotations in GOAT, in particular \emph{Let Ring} labels shown in Figure~\ref{fig:example}, which were identified through manual spectral analysis of the audio. This suggests that some apparent errors may reflect annotation gaps that mislead the training process rather than true model limitations. Finally, we will investigate whether extending the audio input window length improves accuracy, as preliminary experiments suggest.


\clearpage
\section{Ethics Statement}

This work presents a system for automatic guitar tablature transcription, a tool designed to assist musicians, educators, and researchers in music analysis and learning. The training data used in this study—GOAT and GuitarSet—are publicly available research datasets released under appropriate licenses for academic use. Our system is intended as a transcription aid rather than a generative tool, and does not produce new musical content. We acknowledge that automated transcription systems may have implications for the music industry, particularly regarding the economic value of human transcription work, and encourage responsible deployment in contexts that complement rather than displace human musicianship.

\section{Acknowledgements}
This work is supported by the French National Research Agency, in the framework of the project TABASCO (ANR-22-CE38-0001), and the French Ministry of Culture.

\bibliographystyle{authordate1}
\bibliography{ISMIRtemplate}  

\end{document}